
\magnification=\magstep1
\hoffset=-0.5truecm\voffset=0.3truecm
\nopagenumbers\parindent=0pt
\footline={\ifnum\pageno<1 \hss\thinspace\hss
    \else\hss\folio\hss \fi}
\pageno=-1

\newdimen\windowhsize \windowhsize=13.1truecm
\newdimen\windowvsize \windowvsize=6.6truecm

\def\heading#1{
    \vskip0pt plus6\baselineskip\penalty-250\vskip0pt plus-6\baselineskip
    \vskip2\baselineskip\vskip 0pt plus 3pt minus 3pt
    \centerline{\bf#1}
    \global\count11=0\nobreak\vskip\baselineskip}
\count10=0
\def\section#1{
    \vskip0pt plus6\baselineskip\penalty-250\vskip0pt plus-6\baselineskip
    \vskip2\baselineskip plus 3pt minus 3pt
    \global\advance\count10 by 1
    \centerline{\expandafter{\number\count10}.\ \bf{#1}}
    \global\count11=0\nobreak\vskip\baselineskip}
\def\subsection#1{
    \vskip0pt plus3\baselineskip\penalty-200\vskip0pt plus-3\baselineskip
    \vskip1\baselineskip plus 3pt minus 3pt
    \global\advance\count11 by 1
    \centerline{{\it {\number\count10}.{\number\count11}\/})\ \it #1}}
\def\firstsubsection#1{
    \vskip0pt plus3\baselineskip\penalty-200\vskip0pt plus-3\baselineskip
    \vskip 0pt plus 3pt minus 3pt
    \global\advance\count11 by 1
    \centerline{{\it {\number\count10}.{\number\count11}\/})\ \it #1}}

\def\eol{\hfil\break}
\def\affl#1{\noindent\llap{$^{#1}$}}

\def\({\left(} \def\){\right)}
\def\th{\theta} \def\rhf{\rho_F}
\def\unvec#1#2{\hat #1_{#2}}
\def\ux{\unvec{u}{x}} \def\uy{\unvec{u}{y}} \def\uz{\unvec{u}{z}}
\def\ws{\unvec{w}{s}} \def\wp{\unvec{w}{p}} \def\wt{\unvec{w}{t}}
\def\wq{\unvec{w}{q}}
 \def\uj{\unvec{u}{j}} 
\def\wi{\unvec{w}{i}} \def\wj{\unvec{w}{j}} 
\def\ricci{\epsilon_{ijk}}
\def\xyz{x,y,z} \def\spt{s,p,t} \def\wspt{\ws,\wp,\wt}
\def\xyzset{\{\xyz\}} \def\sptset{\{\spt\}} \def\wsptset{\{\wspt\}}
\def\rone{\unvec{r}{1}} \def\rtwo{\unvec{r}{2}}
\def\aone{\alpha_1} \def\atwo{\alpha_2}
\def\Dij{D_{ij}} \def\Dijs{\Dij^{(s)}} \def\Dijt{\Dij^{(t)}}
\def\Dy{D_y} \def\Dz{D_z} \def\ny{n_y} \def\DY#1{D_{y#1}} \def\RR#1{r_{#1 0}}
\def\vx{v_x} \def\vy{v_y}

\def\d{\,{\rm d}}

\def\pdsh#1{\partial_{#1}}
\def\pdy{\pdsh{y}} \def\pdz{\pdsh{z}} \def\pdth{\pdsh{\th}}
\def\pds{\pdsh{s}} \def\pdt{\pdsh{t}}

\def\xmx#1{\(x-x_{#1}\)}
\def\yovxs{\(y\over x_s\)} \def\O{\hbox{$\cal O$}}

{
\def\cl#1{\hbox to \windowhsize{\hfill#1\hfill}}
\hbox to\hsize{\hfill\hbox{\vbox to\windowvsize{\vfill
\bf
\cl{ MODELLING NON-AXISYMMETRIC BOW SHOCKS}
\bigskip
\cl{Rino~Bandiera$^1$}
\bigskip\rm
\cl{Preprint n.~15/93}

\vfill}}\hfill}}

\vskip5truecm
{\leftskip1.7truecm
\affl{1}Osservatorio Astrofisico di Arcetri,
\eol
Largo E.~Fermi 5, I-50125 Firenze (Italy)

\vfill
Astronomy and Astrophysics, in press
\vglue3truecm
}
\eject
\vglue\windowvsize plus 0pt minus \vsize
\heading{ABSTRACT}

I present an extension of the thin layer approximation to non-axisymmetric bow
shocks.
By choosing a suitable set of curvilinear coordinates that matches the geometry
of a generally distorted bow shock surface, I derive the fluid equations for
the flow.
Analytical expansions are given for the matter flow near the bow shock
stagnation point.
Numerical solutions are also obtained, by using a revised Runge-Kutta method.
This numerical method results to be computationally stable and fast.
When two winds interact, axisymmetric bow shocks are expected to form only if
both winds are spherically symmetric. But stellar winds often present strong
anisotropies: for all these cases a non-axisymmetric model of bow shock is
required.

\medskip
\noindent\underbar{\strut Subject}\ \underbar{\strut headings}:
  hydrodynamics --
  methods: numerical --
  binaries: general --
  ISM: kinematics and dynamics

\vglue\windowvsize plus 1fill minus \vsize
\eject
\pageno=1
\section{Introduction}

The interaction between two winds or between a wind and a moving ambient
medium will usually result in the formation of a bow shock.
A popular astrophysical example of such a problem is, on a planetary scale,
the flow structure in the cometary ionosphere, interacting with the solar wind
(e.g.\ Houpis \& Mendis 1980).

A similar situation is often encountered also on a stellar scale.
I shall point out here some of the various fields of application.
The collision between stellar winds may give effects in symbiotic stars (Girard
\& Willson 1987), as well as in hot binary stars (Kallrath 1991a, 1991b).
Comet shaped regions may also appear when a mass losing star is moving through
the ambient medium.
Ultracompact $H\,II$ regions have been explained as bow shocks formed by
massive stars moving through molecular clouds (Van Buren et al.\ 1990, Mac Low
et al.\ 1991).
Some arcuate features revealed by IRAS close to hot stars (de Vries 1985, Van
Buren \& McCray 1988) may be also explained in terms of bow shocks.
A very faint bow shock has been observed close to the binary millisecond
pulsar PSR~1957+20 (Kulkarni \& Hester 1988, Aldcroft, Romani \& Cordes 1992)
and has been explained as due to the interaction of the relativistic pulsar
wind with neutral hydrogen in the interstellar medium.
Some ring nebulae associated to runaway Wolf-Rayet stars present shapes and
kinematics that are consistent with more or less developed bow shocks.
Good examples are S308, M1-67, NGC~6888, NGC~3199: a detailed study of the
kinematics of M1-67 (Solf \& Carsenty 1982) indicates that the wind-blown
bubble forming the nebula has been considerably distorted in the direction of
the stellar motion by the interaction with the ambient medium.

A bow shock model has been proposed for explaining the structure of the
circumstellar medium around the progenitor of Kepler's SN (Bandiera 1987,
Borkowski et al.\ 1992).
According to that model the optically emitting knots are condensations lying on
a bow shock, that have been recently reached by the supernova blast wave.
However in this case the observed geometry shows a distortion that cannot be
accounted for in detail by using a simple axisymmetric bow shock model.
A distorted bow shock is actually expected, since the progenitor's wind in its
red supergiant phase was likely to be anisotropic.
A study of bow shock asymmetries is feasible for most of the fields mentioned
above, whenever detailed information on the structure and kinematics is
available with good enough spatial resolution.
The principal aim of this kind of investigation is to get information on the
degree of anisotropy of the stellar wind velocity and flux.

In this paper I present an efficient way to compute asymmetric bow shocks.
Since the algorithm is computationally fast it can be used for a model matching
of
observations.
The plan of the paper is as follows. Sect.~2 is devoted to specify the main
assumptions, in comparison with those used by Huang \& Weigert (1982a).
Sect.~3 contains a formal derivations of the fluid equations.
In Sect.~4 I present some analytical solutions and expansions, while in
Sect.~5,
after describing the numerical method used, I comment some numerical solutions.
Sect.~6 concludes.

\section{Assumptions}
The hydrodynamics of bow shocks is generally rather complex, and various sorts
of approximations have been applied by various authors.
I shall consider here a steady state bow shock, formed by the interaction of
two winds, under the assumption that its thickness is negligible in comparison
with other scale lengths.
This assumption is the so-called ``thin layer approximation'', and is
appropriate if the shocked gas can radiate efficiently its internal energy, so
that it shrinks into a thin layer.
Consistently, pressure terms in the hydrodynamic equations will be neglected.
Let us further assume that on any given point of the surface all fluid elements
move with a well defined velocity, tangential to the surface itself.
This also implies a complete mixing of the shocked material from the two winds.
In this case the problem can be reduced from 3-D to 2-D by deriving a set of
fluid equations on a manifold matching the bow shock surface.
In the following I shall further assume that the bow shock solution is
stationary in an inertial reference frame.
This is appropriate for a star moving in an ambient medium but is not exactly
true, for instance, if the bow shock is formed by the interaction between winds
of orbiting stars.
Anyway inertial forces give small effects, provided that the stellar wind
velocities are larger than the orbital velocities.

In this paper I shall conform to the conventions and notation introduced by
Huang \& Weigert (1982a), and I shall extend those conventions when needed.
Huang \& Weigert used a cylindrical coordinates system whose axis crosses the
two wind sources, assumed rotational symmetry about this axis (named $x$-axis),
and defined as $y$-axis an arbitrary direction perpendicular to it.
Our main improvement with respect to Huang and Weigert consists in releasing
the condition of axial symmetry: therefore one must also introduce an azimuthal
angle $\th$.
The bow shock surface can be represented by the equation $x=x(y,\th)$.
This equation is single-valued provided that $\pdy x$ is always not singular.
Such a condition is fulfilled in bow shock geometries, as it will be
shown later on.
In the following, while dealing with space derivatives, I shall substitute the
variable $\th$ with $z$, defined by $\d z=y\d\th$.
Consistently, the unit vectors along cylindrical axes will be called $\ux$,
$\uy$ and $\uz$.

\section{The fluid equations}

\firstsubsection{Divergence terms in curvilinear coordinates}

I shall use a suitable curvilinear coordinates system $\sptset$, in order to
express the fluid equations in a simple form.
This coordinates system is defined by means of its unit vectors $\wsptset$,
where $\ws$ is taken along the fluid motion, $\wp$ is perpendicular to the
surface,
while $\wt$ is orthogonal to the previous two directions.
In these coordinates, the bow shock surface is defined as a surface at constant
$p$, while the fluid velocity is simply $v\ws$.

Let us consider a transformation from cylindrical to curvilinear coordinates
defined by a unitary matrix $C_{ij}$: therefore $C_{ij}=\wi\cdot\uj$, where
$i\in\sptset$ and $j\in\xyzset$.
Let us define the matrices $\Dijs=\wi\cdot\pds\wj$ and $\Dijt=\wi\cdot\pdt\wj$,
where $i,j\in\sptset$.
Since $\{\wi\}$ is an orthonormal basis, any generic matrix
$\Dij=\wi\cdot\partial\wj$ is antisymmetric and can be therefore associated to
a vector $\vec d$ by the relation $\Dij=\sum_k\ricci d_k$, where $\ricci$ is
the alternating tensor.
More specifically, I shall call $\vec s$ and $\vec t$ the vectors associated
with $\Dijs$ and $\Dijt$, respectively.

The matrices $\Dij$ are expressed in terms of the derivatives $\pds$ and
$\pdt$,
that in turn can be derived from $\pdy$ and $\pdz$.
Therefore the matrices $\Dij$ are as follows:
$$\pmatrix{\Dijs\cr\Dijt\cr}=\pmatrix{\ws\cdot\uy&\ws\cdot\uz\cr
\wt\cdot\uy&\wt\cdot\uz\cr}\cdot\pmatrix{\wi\cdot\pdy\wj\cr
\wi\cdot\pdz\wj\cr},\eqno(1)$$
where $\wi\cdot\pdy\wj$ and $\wi\cdot\pdz\wj$ are, in terms of $C_{ij}$ and its
derivatives:
$$\eqalignno{
&\wi\cdot\pdy\wj=\sum_k C_{ik}\pdy C_{jk},&(2a)\cr
&\wi\cdot\pdz\wj=\sum_k C_{ik}\pdz C_{jk}+\(C_{iz}C_{jy}-C_{iy}C_{jz}\)/y.&(2b)
\cr}$$
In deriving the latter formula I explicitely took into account those
derivatives of the basis vectors that in cylindrical coordinates are not
vanishing, namely $\pdth\uy=\uz$ and $\pdth\uz=-\uy$.

As we shall see in the following when writing the fluid equations, some of the
quantities defined above will enter in the formulas for the divergence terms.
In fact, the nabla operator evaluated on the bow shock surface is:
$$\vec\nabla=\ws\pds+\wt\pdt,\eqno(3)$$
Therefore, with the help of previous definitions, we can transform the
divergence terms as follows:
$$\eqalignno{
&\vec\nabla\cdot\(\rhf\vec v\)=\pds\(\rhf v\)+\rhf vt_p,&(4a)\cr
&\vec\nabla\cdot\(\rhf\vec v\vec v\)=\(\pds\(\rhf v^2\)+\rhf v^2t_p\)\ws-\rhf
v^2s_t\wp&\cr
&\qquad+\rhf v^2s_p\wt,&(4b)\cr}$$
where $\rhf$ is the surface density.
Moreover, on the basis of previous definitions:
$$s_p=\wt\cdot\pds\ws,\quad s_t=\ws\pds\wp,\quad t_p=\wt\pdt\ws.\eqno(5)$$
Although the divergence terms are formally simpler in curvilinear coordinates,
in order to solve the fluid equations and to get $x(y,\th)$, $\rhf(y,\th)$ and
$\vec v(y,\th)$ it is convenient to express them in cylindrical coordinates.
This will be done in Sect.~3.3.

\subsection{Source terms in fluid equations}

Let us assume that the bow shock is fed by steady winds that originate from two
sources, labelled $(1)$ and $(2)$.
The $x$-axis is defined as passing through these two sources, and is oriented
in such a way that the $x$ coordinate of source $(2)$ be larger than that of
source $(1)$.
In their papers on the subject Huang \& Weigert (1982a,b) considered the two
following cases: ($x_1=0$, $x_2$ positive), and ($x_1=-\infty$, $x_2=0$)
respectively.
Each wind moves radially from the respective source, providing both mass and
momentum to the bow shock.

Let $W_1$ and $W_2$ be the velocities of the two winds, while $J_1$ and $J_2$
be the respective momentum fluxes ($J=\rho W^2$).
One can define $J=\dot mW/r^2$, where $\dot m$ is the mass loss per steradians
($\dot M=\int\dot m\d\Omega$).
Therefore $J_1$ and $J_2$ scale as the inverse square of the respective radial
distances, $r_1=\sqrt{\xmx1^2+y^2}$ and $r_2=\sqrt{\xmx2^2+y^2}$.
Unlike Huang \& Weigert (1982a,b) now the winds are not constrained to be
isotropic: let $W_1$ and $\dot m_1$ be arbitrary functions of $\th$ and
$\xi_1=\arccos(\xmx1/r_1)$, as well as $W_2$ and $\dot m_2$ be functions of
$\th$ and $\xi_2=\arccos(\xmx2/r_2)$.
The fluid equations can therefore be written as:
$$\eqalignno{
&\vec\nabla\cdot\(\rhf\vec v\)=-(J_1/W_1)(\rone\cdot\wp)+(J_2/W_2)(\rtwo\cdot
\wp),&(6a)\cr
&\vec\nabla\cdot\(\rhf\vec v\vec v\)=-J_1(\rone\cdot\wp)\rone+J_2(\rtwo\cdot
\wp)\rtwo.&(6b)\cr}$$
For convention $\wp$ is oriented towards the region of space that contains
source $(1)$.
The vectors $\rone$ and $\rtwo$ are radial unit vectors from $(1)$ and $(2)$,
respectively.
Their components in cylindrical coordinates are:
$$\eqalignno{
&\rone=(\xmx1/r_1,\;y/r_1,\;0),&(7a)\cr
&\rtwo=(\xmx2/r_2,\;y/r_2,\;0).&(7b)\cr}$$
For a source $(1)$ located at $x_1=-\infty$ $\rone$ approaches $\ux$.

\subsection{Explicit form of the fluid equations}

Although I did not assume axial symmetry the condition that the two wind
sources are located along the $x$-axis implies that each flux line lies on a
plane containing the $x$-axis, namely at constant $\th$.  In fact the component
along $\uz$ of the external input of momentum is zero.
On the other hand if thermal pressure is negligible there is no way for the
flow to self-generate a velocity component along $\uz$.
Another consequence of this result is that at the intersection of the bow shock
with the $x$-axis the fluid velocity vanishes: therefore a stagnation point
must be present there.
It is also worth noticing that the stagnation point is unique.
In fact the $y$ component of the momentum input provided by the winds cannot be
negative: therefore, as soon as the flux gets a positive $y$ velocity component
it cannot be stopped anymore.
Incidentally, this also proves that the equation $x=x(y,\th)$ defining the bow
shock surface is single-valued, as assumed at the beginning of this section.
All these results will allow us to noticeably simplify the modelling of the
flow.

For instance, in the $\xyzset$ coordinates system the fluid velocity can be
generally expressed as $\vec v=\vx\ux+\vy\uy$.
The condition that the velocity is tangential to the surface gives
$\vx=\Dy\vy$.
The velocity components can then be written as $\vx=\Dy v/\ny$ and $\vy=v/\ny$.
In order to simplify the notation here and in the following I shall use $\Dy$
as a shortcut for $\pdy x$ and $\Dz$ for $\pdz x$, and well as the quantities
$\ny^2=1+\Dy^2$ and $n^2=1+\Dy^2+\Dz^2$.

The explicit form of the matrix that describes the coordinates transformation
is:
$$C_{ij}=\pmatrix{\Dy/\ny&1/\ny&0\cr-1/n&\Dy/n&\Dz/n\cr\Dz/n\ny&-\Dy\Dz/n\ny&
\ny/n\cr}.\eqno(8)$$
In the definition of the coordinate system $\sptset$ I used the following
prescriptions on the orientations: the $x$ component of $\wp$ is negative,
while $\wt$ has been taken such as $\sptset$ have the same chirality of
$\xyzset$ set of coordinates.

By introducing the explicit form of $C_{ij}$ in Eqs.\ 1 and 2 after some
algebra one can evaluate the following quantities:
$$\eqalignno{
&s_p=\Dz\pdy\Dy/n\ny^3,&(9a)\cr&s_t=\pdy\Dy/n\ny^2,&(9b)\cr
&t_p=\(-\Dy\Dz^2\pdy\Dy+\ny^2\Dz\pdz\Dy+\ny^4/y\)/n^2\ny^3,&(9c)\cr}$$
that must be substituted into Eqs.\ 4; also $\pds$ simply translates into
$(1/\ny)\pdy$.
One can show that, as expected, the projection of Eq.~4b along $\uz$ vanishes.
Therefore from this vectorial equation one can extract only 2 scalar
expressions
which are linearly independent: for convenience I shall choose the directions
of $\ws$ and of $\wq=(-\ux+\Dy\uy)/\ny=(\ny\wp-\Dz\wt)/n$ ($\wq$ has been
defined as orthogonal to $\ws$ and to $\uz$).
{}From Eqs.\ 4 one can derive:
$$\eqalignno{
&\vec\nabla\cdot\(\rhf\vec v\)=\(N/y\ny\)\pdy\(y\Phi/vN\),&(10a)\cr
&\vec\nabla\cdot\(\rhf\vec v\vec
v\)\cdot\ws=\(N/y\ny\)\pdy\(y\Phi/N\),&(10b)\cr
&\vec\nabla\cdot\(\rhf\vec v\vec v\)\cdot\wq=-\Phi/R,&(10c)\cr}$$
where $\Phi=\rhf v^2$, $N=\ny/n$ and $R=\ny^3/\pdy\Dy$: $R$ indicates the
curvature radius in the trajectories.
I also used the identity $\pdz\Dy=\pdy\Dz+\Dz/y$, equivalent to
$\pdsh{\th y}x=\pdsh{y\th}x$.
The basic difference with respect to the axially symmetric case is the presence
of the term $N$, corresponding to $\wq\cdot\wp$, that in the symmetric case is
equal to unity.

Also the projections of the source terms along $\ws$ and $\wq$ allow us some
further simplification.
Again following the notation of Huang \& Weigert (1982a) one can introduce the
angles $\aone$ and $\atwo$, that can be connected to our notation as follows:
$$\eqalignno{
&\cos\aone=\quad\rone\cdot\ws=\(\xmx1\Dy+y\)/r_1\ny,&(11a)\cr
&\sin\aone=    -\rone\cdot\wq=\(\xmx1-y\Dy\)/r_1\ny,&(11b)\cr
&\cos\atwo=\quad\rtwo\cdot\ws=\(\xmx2\Dy+y\)/r_2\ny,&(11c)\cr
&\sin\atwo=\quad\rtwo\cdot\wq=\(y\Dy-\xmx2\)/r_2\ny.&(11d)\cr}$$
In this way one can complement Eqs.\ 10 with the corresponding source terms and
obtain:
$$\eqalignno{
&\pdy(y\Phi/vN)/y\ny=(J_1/W_1)\sin\aone+(J_2/W_2)\sin\atwo,&(12a)\cr
&\pdy(y\Phi/N)/y\ny=J_1\sin\aone\cos\aone+J_2\sin\atwo\cos\atwo,&(12b)\cr
&(\Phi\pdy\Dy)/\ny^3N^2=J_1\sin\aone^2-J_2\sin\atwo^2,&(12c)\cr}$$
where I used the identity
$\unvec{r}{i}\cdot\wp=\(\unvec{r}{i}\cdot\wq\)\(\wq\cdot\wp\)$.
An interesting property of the last two equations is that they do not contain
neither the flux velocity $v$, nor the velocities of the input winds ($W_1$ and
$W_2$).
Therefore, as in the symmetric case, also here a solution for the bow shock
shape as well as for the momentum flux ($\Phi$) is affected by $J_1$ and $J_2$,
but not by $W_1$ and $W_2$.
For a source (1) at $-\infty$, representing a parallel flow, one should take
$J_1=\rho_1W_1^2$ (where $\rho_1$ is the wind density), $\cos\aone=\Dy/\ny$ and
$\sin\aone=1/\ny$.

\section{Analytical solutions}

\firstsubsection{Isotropic winds}

As an exercise one can derive the fluid equations for a bow shock formed by the
interaction of two otherwise isotropic winds.
In this case the solution is expected to be axially symmetric.

The special case with axial symmetry is readily obtained by taking $N=1$ in
Eqs.\ 12.
Incidentally it is worth noticing that the analog in Huang \& Weigert (1982a,
b) of our Eq.~12b was incorrect, since there the geometrical divergence term
was missing.

In the isotropic case Eq.\ 12a can be integrated analytically, giving:
$$\eqalignno{
&y\rhf v=\(\dot M_1/4\pi\)\(1-\xmx1/r_1\)&\cr
&\qquad+\(\dot M_2/4\pi\)\(1+\xmx2/r_2\).&(13)\cr}$$
For a source (1) at $-\infty$ and a source (2) at the origin (Huang \&
Weigert 1982b), this equation reads:
$$y\rhf v=\(\rho_1W_1\)y^2/2+\(\dot M_2/4\pi\)\(1+x/r_2\).\eqno(14)$$

Solutions for $x$ and $\Phi$ can be written more easily in terms of the
following parameter:
$$j=\sqrt{\dot M_1W_1/\dot M_2W_2}.\eqno(15)$$
The position of the bow shock stagnation point is $x_a=(x_1+jx_2)/(j+1)$, while
the distance of the stagnation point from source (2), used hereafter as a scale
length, is $x_s=x_2-x_a=(x_2-x_1)/(j+1)$.
A convenient scale for $\Phi$ is $\Phi_s=\dot M_2W_2/4\pi x_s$.
The quantities $x$ and $\Phi$ can be evaluated from equations 12b and 12c as
polynomial expansions in $y$:
$$\eqalignno{
&{x-x_a\over x_s}={3(j-1)\over10j}\yovxs^2+{3(j-1)(5j^2-42j+5)\over1400j^3}
\yovxs^4\cr
&\qquad+\O\yovxs^6,&(16a)\cr
&{\Phi\over\Phi_s}={(j+1)\over3j}\yovxs^2-{(j+1)(j^2+8j+1)\over25j^3}\yovxs^4&
  \cr
&\qquad+\O\yovxs^6.&(16b)\cr}$$
In the limiting case when source (1) is at $-\infty$ (and then $j=\infty$), we
have:
$$x_s=\sqrt{\dot M_2W_2/4\pi\rho_1W_1^2}.\eqno(17)$$
The scaling for $\Phi$ can be alternatively written as $\Phi_s=\rho_1W_1^2x_s$.
Using these relations, also in the limit $j=\infty$ the solutions for $x$ and
$\Phi$ are simply obtained from Eqs.\ 16.

\subsection{Behaviour near the stagnation point}

In this section I shall study the behaviour of the flow near the stagnation
point, in the generic case of anisotropic sources, by expanding the relevant
quantities in Taylor series about $y=0$.
Let us first consider the expansion to the second order of the quantity $x$,
and consistently of other geometrical quantities:
$$\eqalignno{
&x=x_0+\DY0y+\DY1y^2/2+\O\(y^3\),&(18a)\cr
&\Dy=\DY0+\DY1y+\O\(y^2\),&(18b)\cr
&r_1=\RR1+\DY0y+\(1+\DY1\)y^2/2\RR1+\O\(y^3\),&(18c)\cr
&r_2=\RR2+\DY0y+\(1-\DY1\)y^2/2\RR2+\O\(y^3\),&(18d)\cr}$$
where the coefficients of the expansion are generic functions of $\th$.
Let $Y=\dot mW=Jr^2$ be the momentum loss per steradian, and let us expand the
momentum losses from the two sources near the direction to the bow shock
stagnation point:
$$\eqalignno{
&Y_1=Y_{10}+Y_{11}\(y/r_1\)+Y_{12}\(y/r_1\)^2+\O\(y^3\),&(19a)\cr
&Y_2=Y_{20}+Y_{21}\(y/r_2\)+Y_{22}\(y/r_2\)^2+\O\(y^3\).&(19b)\cr}$$
These expansions can be reduced to pure polynomial expansions in $y$ by the
use of Eqs.\ 18c, 18d.

Our aim is to derive series expansions for Eqs.\ 12.
One may show that also in the non-axisymmetric case the quantities $\Phi$
and $v$ can be expanded as follows:
$$\Phi=\Phi_0y^2+\O\(y^3\),\qquad v=v_0y+\O\(y^2\).\eqno(20)$$
These expansions have been used in the evaluation of the left sides of Eqs.\
12.
The right sides, instead, are evaluated using the definitions in Eqs.\ 11.
The simplest Taylor expansions of Eqs.\ 12 containing non-trivial terms on
both sides of the equations are respectively:
$$\eqalignno{
&2\Phi_0/v_0Nn_y=\(\({Y_{10}\over\RR1^2W_{10}}+{Y_{20}\over\RR2^2W_{20}}\)
  +\O\(y\)\)\Big/n_y,&(21a)\cr
&\(3\Phi_0/Nn_y\)y=\bigg(\DY0\({Y_{10}\over\RR1^2}-{Y_{20}\over\RR2^2}\)&\cr
&\qquad+\bigg(\(1-3\DY0^2\)\({Y_{10}\over\RR1^3}+{Y_{20}\over\RR2^3}\)&\cr
&\qquad+\DY0\({Y_{11}\over\RR1^3}-{Y_{21}\over\RR2^3}\)\bigg)y+\O\(y^2\)\bigg)
  \Big/n_y^2,  &(21b)\cr
&\(\Phi_0\DY1/N^2n_y^3\)y^2=\bigg(\({Y_{10}\over\RR1^2}-{Y_{20}\over\RR2^2}\)
  +\bigg(\({Y_{11}\over\RR1^3}-{Y_{21}\over\RR2^3}\)  &\cr
&\qquad
  -4\DY0\({Y_{10}\over\RR1^3}+{Y_{20}\over\RR2^3}\)\bigg)y
  +\bigg(\({Y_{12}\over\RR1^4}-{Y_{22}\over\RR2^4}\)  &\cr
&\qquad
  -10\DY0\({Y_{11}\over\RR1^4}+{Y_{21}\over\RR2^4}\)
  +4\(5\DY0-1\)\({Y_{10}\over\RR1^4}-{Y_{20}\over\RR2^4}\)  &\cr
&\qquad
  -6\DY1\({Y_{10}\over\RR1^3}+{Y_{20}\over\RR2^3}\)\bigg){y^2\over2}
  +\O\(y^3\)\bigg)\Big/n_y^2.&(21c)\cr }$$
Here $W_{10}$ and $W_{20}$ indicate the velocities of the two winds along the
$x$ axis.

In Eqs.\ 21b, 21c the right sides formally contain terms with $y$ powers lower
than in the respective left sides.
These equations are self-consistent only if the coefficients of those terms are
vanishing, namely if:
$$\eqalignno{
&Y_{10}/\RR1^2=Y_{20}/\RR2^2\qquad\(=J_0\),&(22a)\cr
&\DY0=\(Y_{11}/\RR1^3-Y_{21}/\RR2^3\)/4J_0\(1/\RR1+1/\RR2\).&(22b)\cr }$$
$J_0$ represents the momentum fluxes of the two winds at the stagnation point:
as expected, they balance each other.
The remaining terms can be used to evaluate the other quantities:
$$\eqalignno{
&\Phi_0=\(N\over3n_y\)\bigg(\(1-3\DY0^2\)J_0\({1\over\RR1}+{1\over\RR2}\)&\cr
&\qquad
  +\DY0\({Y_{11}\over\RR1^3}-{Y_{21}\over\RR2^3}\)\bigg),&(23a)\cr
&v_0=2\Phi_0/J_0\({1\over W_{10}}+{1\over W_{20}}\)N,&(23b)\cr
&\DY1=-{3\over2}Nn_y^2\bigg(4\(1-5\DY0^2\)J_0\({1\over\RR1^2}-{1\over\RR2^2}\)
  &\cr
&\qquad
  +10\DY0\({Y_{11}\over\RR1^4}+{Y_{21}\over\RR2^4}\)
  -\({Y_{12}\over\RR1^4}-{Y_{22}\over\RR2^4}\)\bigg)\Big/  &\cr
&\qquad
  \bigg(\(9Nn_y^2+1-3\DY0^2\)J_0\({1\over\RR1}+{1\over\RR2}\)  &\cr
&\qquad
  +\DY0\({Y_{11}\over\RR1^3}-{Y_{21}\over\RR2^3}\)\bigg).&(23c) }$$
The interaction of a wind with a parallel flow is represented by simply taking
$\RR1=\infty$.

In the isotropic case ($Y_{11}=Y_{21}=Y_{12}=Y_{22}=0$) one can easily obtain
the expansion given in Eqs.\ 16 (in this case limited to $y^2$) just using the
following correspondences between notations here and there:
$$\RR1=jx_s,\qquad\RR2=x_s,\qquad J_0=\Phi_sx_s.\eqno(24)$$
The expansion near the stagnation point has been used as a starting point in
numerical computations, as described in the next section.

\section{Numerical Solutions}

\firstsubsection{Outline of the numerical technique}

I have shown that by the thin layer approximation the intrinsically 3-D
problem of an asymmetric bow shock can be reduced to 2-D.
Now I shall describe how it can be further reduced, for the numerical
modelling, in a series of parallel 1-D problems, thus requiring only the
solution of ordinary differential equations.

Since no pressure forces are present, the flows at different azimuthal angles
are, in a physical sense, mutually independent.
However as a consequence of the thin layer approximation a fictious dependence
is contained in the quantity $N$, under the form of azimuthal derivatives of
$x$.

Eqs.\ 12 can be seen as a set of partial differential equations in the four
variables $x$, $\Dy$, $\Phi$ and $v$ (the missing equation is simply
$\pdy x=\Dy$); the azimuthal derivatives are however hidden in the quantity
$N$, containing the term $\Dz$.
Let us take a cylindrical grid with $N_\th$ azimuthal bins, and define a vector
$V$ containing for any given $y$ the four variables at all values of $\th$: the
dimension of this vector is then $4N_\th$.
Given $x$, also $\Dz$ can be computed by numerical differentiation; namely each
$\Dz$ is well approximated by a linear combination of $x$s at different values
of $\th$.
Therefore all derivatives with respect to $y$ can be expressed as functions
of $\Dy$, $\Phi$ and $v$ at the same $\th$, but also of $x$ at various nearby
azimuthal positions.
In a compact vectorial notation, one can then express Eqs.\ 12 as
$\partial_yV=F(V)$, and then use a routine for integration of a set of
differential equations.

Using $y=0$ as a starting point leads to serious numerical problems, because
at the stagnation point various quantities vanish at the same time.
For this reason one can get the starting behaviour from the Taylor expansion
computed in the previous section.
With this expedient the equations can be easily computed with a standard
Runge-Kutta algorithm, presenting a good convergence for reasonable choices of
winds.

\subsection{Examples of numerical models}

In this section I present some numerical results.
In order to reduce the number of parameters, I shall consider here only the
interaction of a wind with a parallel flow ($x_1=-\infty$).
However the qualitative behaviour of the results is not much different from the
case with two centered winds.

I also model the anisotropy in momentum flux from source (2) by a quadrupolar
term added to the isotropic part; namely:
$$Y_2=Y_*\(1+\Delta_Y\(1-3\cos^2(\chi)\)/2\).\eqno(24)$$
This form presents axial symmetry, as expected of a stellar wind, but with
respect to an axis that is arbitrarily oriented in space.
In the following examples I shall take a wind velocity independent of the
direction, even though the code can adequately account also for velocity
anisotropy.
Anyway in Sect.~3.2 it has been shown that the wind velocities will affect
only the gas velocity
along the bow shock, but will not affect neither the bow shock shape nor the
momentum flux.

Let $\chi$ be the angle between a given direction and the direction of that
axis of symmetry.
If in a reference frame centered on source (2) the direction of the symmetry
axis is represented by the angles $(\xi_*,\th_*)$, the angle $\chi$
relative to a generic direction $(\xi_2,\th)$ is obtained by:
$$\cos\(\chi\)=\cos\(\xi_2\)\cos\(\xi_*\)+\sin\(\xi_2\)\sin\(\xi_*\)
  \cos\(\th-\th_*\).\eqno(25)$$
The quantity $\Delta_Y$ represents the amount of anisotropy in the wind
momentum flux.
It must be bound to the interval $[-2,+1]$: negative values indicate an
excess along the symmetry (polar) axis, while positive values indicate an
excess on the orthogonal (equatorial) plane.

The results are summarized in a few figures.
Fig.~1 shows the effects of the wind anisotropy on the bow shock shape.
Four frames are presented, with $\Delta_Y$ ranging from -2 to 1 and a
constant angle $\xi_*=30^\circ$ between the direction of the parallel flow and
that of the wind symmetry axis.
Each frame contains the shape of the bow shock (the star is indicated by a
black spot), and on its right the corresponding angular dependence of the
momentum flux of the related wind.
All solutions are limited here to $y<3$.
In order to make the 3-D view more apparent, I used for all figures a tilt of
$30^\circ$ about the horizontal axis (at $x=0$, in our notation).
Moreover the local thickness of the grid should give a feeling of the third
dimension.
Fig.~1c indicates the isotropic case: therefore in this case $\xi_*$ loses its
meaning.
Here and in the following figures I used grids composed of a set of
curves at constant $\th$ (radial lines as seen from the $x$ axis) and a set of
curves at constant $y$ (concentric circles as seen from the $x$ axis).

Fig.~2 is analog to the previous one.
Here however $\Delta_Y$ is fixed to -1, while $\xi_*$ takes the values
$10^\circ, 30^\circ, 50^\circ, 70^\circ,$ respectively.
Let us note that Fig.~1b and Fig.~2b are the same.

Finally Fig.~3 gives an impression of the behaviour of the various physical
quantities.
For this figure I used an hybrid representation.
For any given quantity function of $y$ and $\th$, I originally used the
horizontal plane for space position, while the vertical axis is reserved for
the quantity itself.
For quantities whose dimensions are different from a pure length a suitable
scale has been chosen in order to fill efficiently the plotting area.
Finally, in these hybrid system of coordinates I again applied a rotation of
$30^\circ$.
If the plotted quantity is $x$ (Fig.~3a), the bow shock shape is reproduced.
Since I used here a model with $\xi_*=50^\circ$ and $\Delta_Y=-1$, Fig.~3a
is the same as Fig.~2c.

The remaining frames of Fig.~3 cannot be easily used for a quantitative
discussion.
However one can obtain from them a direct qualitative impression of the
behaviour of the various quantities.
As $x$, also $\Phi$ and $v$ are null at $y=0$.
The surface density $\rho_F$ is instead positive at all points.
Since near the stagnation point $\Phi$ is quadratic with $y$, the ``tip'' of
the surface looks smooth; it is sharp instead for $v$, because it is linear
with $y$.
The kind of asymmetry of the various quantities can be qualitatively
understood.
In fact, as seen from Fig.~2c, this model corresponds to a polar excess of
momentum flux.
On the left side a lobe is pointing almost opposite to the parallel flow.
Therefore on that side $x$ is minimum (maximum protrusion against the external
flow); $\Phi$ and $v$ are, for the same value of $y$, lower than on the right
side, since the parallel flow gives near the stagnation point a negative
contribution to the flow along the bow shock.
Finally, connected to the slower motions on the left side we have there also
the maximum of surface dansity, even larger than that at the stagnation point.

\section{Conclusions}

In this paper I have shown that one can release the assumption of axisymmetry
for the bow shock and still obtain fluid equations which are formally similar
to those in the symmetric case.
Crucial in this approach is the choice of a curvilinear coordinates system that
matches the shape of the bow shock.
Important for the development of this model have also been the steady state
assumption and the thin layer approximation.
The latter limits the quantitative use of the results to the case in which the
cooling time is shorter that the dynamical one; furthermore this approximation
implies a complete mixing between the two flows in the bow shocks.
Another limitation of this model is the use of an inertial frame, that make it
of little use for the case of binary stars with slow winds, compared to
orbital motions.
Matter of a future work will be a generalization of the present work by
releasing some of the assumptions listed above.

\heading{Acknowledgements}

I am grateful to A. Natta for important suggestions, to Y. Chen for
interesting discussions on the argument.
I am indebted to M. Salvati for a careful reading of the manuscript.

\heading{References}
\leftskip12pt\parindent=-12pt \def\ref{\par}

\ref Aldcroft, T.L., Romani, R.W., Cordes, J.M., 1992,
  ApJ 400, 638

\ref Bandiera, R., 1987,
  ApJ 319, 885

\ref Borkowski, K.J., Blondin, J.M., Sarazin, C.L., 1992,
  ApJ 400, 222

\ref de Vries, C.P., 1985,
  A\&A 150, L15

\ref Girard, T., Willson, L.A., 1987,
  A\&A 183, 247

\ref Houpis, H.L.F., Mendis, D.J., 1980,
  ApJ 239, 1107

\ref Huang, R.Q., Weigert, A., 1982a,
  A\&A 112, 281

\ref Huang, R.Q., Weigert, A., 1982b,
  A\&A 116, 348

\ref Kallrath, J., 1991a,
  A\&A 247, 434

\ref Kallrath, J., 1991b,
  MNRAS 248, 653.

\ref Kulkarni, S.R., Hester, J.J., 1988,
  Nature 335, 801

\ref Mac Low, M.-M., Van Buren, D., Wood, D.O.S., Churchwell, E., 1991,
  ApJ 369, 395

\ref Solf, J., Carsenty, U., 1982,
  A\&A 116, 54

\ref Van Buren, D., McCray, R., 1988,
  ApJ 329, L93

\ref Van Buren, D., Mac Low, M.-M., Wood, D.O.S., Churchwell, E., 1990,
  ApJ 353, 570

\leftskip0pt\parindent=0pt

\heading{FIGURE CAPTIONS}
\def\figure#1#2{Figure #1: #2\smallskip}

\figure{1}{Bow shock shapes, for wind anisotropy of constant orientation
($\xi_*=30^\circ$) and varying amount ($\Delta_Y$). On the right side of each
frame is a sketch of the angular dependence of the wind momentum flux}

\figure{2}{Like Fig.~1, but for wind anisotropy of constant amount
($\Delta_Y=-1$) and varying orientations}

\figure{3}{Behaviour of various physical quantities for the case
$\xi_*=50^\circ$ and $\Delta_Y=-1$}

\bye